\documentclass[amsmath, amssymb, aps, superscriptaddress, reprint, pra]{revtex4-2}

\usepackage{graphicx}
\usepackage{dcolumn}
\usepackage{bm}
\usepackage{hyperref}

\usepackage[caption=false]{subfig}
\usepackage{tikz}
\usepackage{placeins}

\begin{document}

\title{Demonstration of transport in an ion trap design for two-dimensional lattices}

\author{M. Pfeifer}
\email[Corresponding author, ]{MichaelDieter.Pfeifer@Infineon.com}
\affiliation{Institut f{\"u}r Experimentalphysik, Universit{\"a}t Innsbruck, Technikerstra{\ss}e 25/4, 6020 Innsbruck, Austria}
\affiliation{Infineon Technologies Austria AG, Siemensstra{\ss}e 2, 9500 Villach, Austria}
\author{M. Valentini}
\affiliation{Institut f{\"u}r Experimentalphysik, Universit{\"a}t Innsbruck, Technikerstra{\ss}e 25/4, 6020 Innsbruck, Austria}
\author{M. Dietl}
\affiliation{Institut f{\"u}r Experimentalphysik, Universit{\"a}t Innsbruck, Technikerstra{\ss}e 25/4, 6020 Innsbruck, Austria}
\affiliation{Infineon Technologies Austria AG, Siemensstra{\ss}e 2, 9500 Villach, Austria}
\author{F. Anmasser}
\affiliation{Institut f{\"u}r Experimentalphysik, Universit{\"a}t Innsbruck, Technikerstra{\ss}e 25/4, 6020 Innsbruck, Austria}
\affiliation{Infineon Technologies Austria AG, Siemensstra{\ss}e 2, 9500 Villach, Austria}
\author{S. Schey}
\affiliation{Infineon Technologies Austria AG, Siemensstra{\ss}e 2, 9500 Villach, Austria}
\affiliation{Department of Physics, Stockholm University, SE-106 91 Stockholm, Sweden}
\author{J. Wahl}
\affiliation{Institut f{\"u}r Experimentalphysik, Universit{\"a}t Innsbruck, Technikerstra{\ss}e 25/4, 6020 Innsbruck, Austria}

\author{P. C. Holz}
\affiliation{Alpine Quantum Technologies GmbH, Technikerstra{\ss}e 17/1, 6020 Innsbruck, Austria}
\author{C. R\"ossler}
\affiliation{Infineon Technologies Austria AG, Siemensstra{\ss}e 2, 9500 Villach, Austria}
\author{Y. Colombe}
\affiliation{Infineon Technologies Austria AG, Siemensstra{\ss}e 2, 9500 Villach, Austria}
\author{P. Schindler}
\affiliation{Institut f{\"u}r Experimentalphysik, Universit{\"a}t Innsbruck, Technikerstra{\ss}e 25/4, 6020 Innsbruck, Austria}

\date{\today}

\begin{abstract}
Microfabricated ion trap chips are at the core of some of the most advanced quantum computers. How a large number of ions is arranged and controlled on an ion trap chip depends on the chosen trap architecture. One such architecture is the quantum spring array (QSA)~\cite{Valentini2025}. In the QSA architecture, ion chains are arranged in a two-dimensional lattice and interact with ion chains in neighboring sites in the radial and axial directions of the respective chain. This interaction, or coupling, is mediated by the Coulomb force while keeping ions in separate trapping sites, and scales inversely with the third power of the separation. The capability to control the distance between ions in the lattice is thus essential. In previous works~\cite{Holz2020, Valentini2025}, the radial separation between ions was tuned by controlling the rf pseudo-potential, which revealed to be experimentally challenging to realize while maintaining low heating rates. In this work, we present an ion trap chip design that allows tuning of the radial distance between ions using only dc voltages. The radial transport is executed between different interaction zones, designated for quantum operations, through specifically designed transition zones. A prototype of this type of ion trap chip was microfabricated on fused silica substrate. Its functionality is characterized by demonstrating dc-controlled radial transport of a single ion through a transition zone and measuring stray fields and ion heating rates in the center of the trap. Moreover, the fabrication of a multi-metal layer version of such a trap is presented as a scaling path for the presented chip design. 
\end{abstract}

\maketitle

\section{\label{sec:introduction}Introduction}
Quantum processors based on ion trap chips are at the core of some of the most advanced quantum computers~\cite{Moses2023, Loeschnauer2024, Smith2025, Bruzewicz2019, Hughes2025, Ransford2025}. In such devices, quantum information is stored in internal states of atomic ions and interaction between ion qubits uses shared motional modes of oscillation~\cite{Cirac-Zoller1995, Schmidt-Kaler2003, Molmer-Sorensen2000} or Rydberg interaction~\cite{Rydberg2020}.

Several modular architectures~\cite{Kielpinski2002, Valentini2025, Holz2020, Kumph2016, Jain2020} have been proposed and implemented to scale up the number of ions controlled by the quantum processors. They differ in the way in which connectivity between subregisters is realized, either by splitting, merging and shuttling of ion chains~\cite{Pino2022, Ransford2025}, nearest-neighbor coupling in an ion lattice~\cite{Valentini2025, Holz2020, Mielenz2016} or photonic interconnects~\cite{Saha2023, Riedel2026}. In the quantum charge-coupled device (QCCD) architecture~\cite{Kielpinski2002}, ions are shuttled between different interaction sites on a grid of tracks intersecting at junctions~\cite{Wright2013, Burton2023}. Ion chains are split and merged~\cite{Kaufmann2014, Palmero2015} so that ions that need to interact are brought to a common potential well. 

Variants of QCCD architectures include the `center to left or right' (C2LR)~\cite{Delaney2024} and the `wiring using integrated switching electronics' (WISE)~\cite{Malinowski2023} schemes. They both operate on a grid of tracks with X-junctions, but differ in the way in which a large number of dc electrodes is wired. 

In the quantum spring array (QSA) architecture~\cite{Valentini2025, Holz2020}, the ions are arranged in chains in a two-dimensional lattice and interact with the nearest neighboring lattice sites. At each lattice site is a subregister consisting of a one-dimensional chain of ions. Full ion-to-ion connectivity within each lattice site can be achieved by individually addressed laser pulses and entanglement within the subregister can be created by using the common mode of motion of the ion chain.

In the QSA architecture, the connectivity between neighboring lattice sites is mediated by the Coulomb interaction between the axial modes of oscillation of the ion chains. The fundamental situation is the coupling of two ions in separate wells of a double-well potential. The corresponding coupling strength is given by~\cite{Brown2011, Harlander2011, Valentini2025}: 
\begin{equation}
    \Omega_{\mathrm{ex}} = \frac{\kappa}{2\pi\varepsilon_{0}} \frac{q_{1}q_{2}}{\sqrt{m_{1}m_{2}}\sqrt{\omega_{1}\omega_{2}}s_{0}^{3}},\label{eq:coupling_rate}
\end{equation}
where $s_{0}$ is the distance between the subregisters, $q_{1}, q_{2}$ their respective charge, $m_{1}, m_{2}$ their respective mass and $\omega_{1}, \omega_{2}$ the axial motional frequencies of the two ions. Equation~\ref{eq:coupling_rate} holds strictly only for point charges, for ion chains the coupling rate does not scale linearly in the charge~\cite{Valentini2025}. The dimensionless pre-factor $\kappa \in \{1, 1/2\}$ depends on whether the interaction is along the axis of the ion chains (axial direction, $\kappa = 1$) or normal to it (radial direction, $\kappa = 1/2$)~\cite{Holz2020}. 

In equation~\ref{eq:coupling_rate}, the coupling strength between lattice sites scales with $\Omega_{\mathrm{ex}} \propto s_{0}^{-3}\omega^{-1}$, with their distance $s_{0}$ and the frequency $\omega$ of the modes (at resonance $\omega_{1} = \omega_{2} = \omega$) used for the interaction. The two parameters that need to be tuned in any concrete implementation of the QSA architecture are therefore $s_{0}$ and $\omega$. The axial mode frequency $\omega$ can be tuned by changing the curvature of the dc potential by application of appropriate dc voltages. 

Doing so does not increase the total confinement, but merely redistributes the existing rf pseudo-potential confinement $\vec{\nabla}^{2}\varphi_{\mathrm{PP}}$~\cite{House2008, Zhang2022, Burton2023}. The pseudo-potential~$\varphi_{\mathrm{PP}}$, created by the rf electrodes~\cite{House2008}, is calculated from the rf electric field $\vec{E}_{\mathrm{rf}} = \vec{E}_{0}\cos{\Omega_{\mathrm{rf}} t}$ by $\varphi_{\mathrm{PP}} = (e^{2}/4m\Omega_{\mathrm{rf}}^{2})|\vec{E}_{0}|^{2}$, where $m$ is the ion mass and $e$ the elementary charge~\cite{House2008}.

For nearest-neighbor ion connectivity in an ion lattice with adjustable coupling strength, the double-well coupling between lattice sites needs to be controllable in the axial and radial directions. In the axial direction, the distance $s_{0}$ between ions in different wells of a double-well potential can be tuned with dc voltages. Adjustment of the axial distance and strong coupling along the axial direction in a double-well potential have been demonstrated in past experiments~\cite{Valentini2025, Mourik2023, Kaushal2020, Walther2012}. 

In the radial direction, the situation is more challenging. In the past, the double-well distance $s_{0}$ has been adjusted by rf shuttling \cite{Holz2020, Valentini2025}. Consider, to fix ideas, an ion trap chip with three long rectangular rf electrodes and thus two rf nulls in a double-well potential. In rf shuttling, one varies the ratio of the rf voltages on the rf electrodes in the trap~\cite{Holz2020, Tanaka2021, Valentini2025} to control the rf pseudo-potential. It turns out, however, that, while rf shuttling has been demonstrated experimentally~\cite{Valentini2025, Tanaka2021}, it could not be demonstrated with heating rates low enough for performing coherent quantum operations. The cause of this heating is still under investigation~\cite{Valentini2025}. Rf shuttling requires precise control of the amplitude and phase of the applied rf voltages and low corresponding noise, which is technically challenging~\cite{Valentini2025}.

\begin{figure}
    \centering
    \includegraphics[width=1\linewidth]{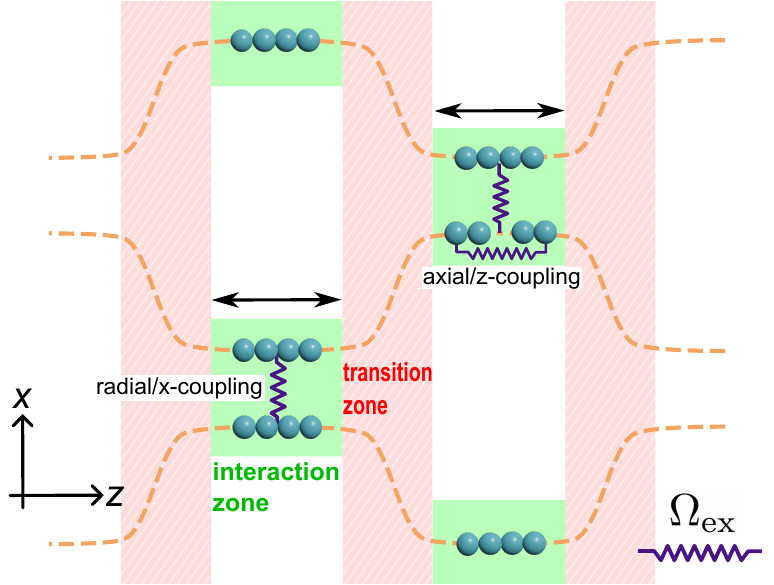}
    \caption{Schematic illustration of the trap geometry investigated in this paper. Rf minimum paths (orange dashed lines) and ion chains (turquoise balls) are forming a two-dimensional lattice for implementing a QSA architecture~\cite{Valentini2025}. The ion chains interact via the Coulomb interaction (purple zig-zag lines) in the ``interaction zones" (green areas) and can be shuttled, using dc electrodes, along the rf minimum paths through ``transition zones" (red areas) between interaction zones. They can interact in the radial direction between lattice sites, as shown on the left of the figure, or axially within a lattice site, as shown on the right.}
    \label{fig:intro}
\end{figure}

In this paper, we present and characterize a demonstrator chip for an implementation of tunable radial coupling that does not require variable rf amplitudes. The solution we propose is to use an ion trap chip design that features curved rf electrodes~\cite{Holz2020-2, Schulz2008}. 
The design is illustrated in figure~\ref{fig:intro}.
The curved rf electrodes create rf minimum paths (orange dashed lines) in the pseudo-potential landscape above the ion trap chip. Unlike the case of linear ion traps with rectangular rf electrodes where the rf minima are straight lines, here the minima follow curved paths. All rf electrodes are driven by the same rf voltage source and are arranged in such a way that the rf minimum paths form a tile pattern, where each path comes close to the adjacent paths above and below in an alternating fashion. 

We call the areas where adjacent rf minimum paths have minimal separation ``interaction zones" and the areas where they grow apart ``transition zones". Ions can be moved between transition zones on the same rf minimum path using dc shuttling. For coupling in the radial direction ($x$-direction), two ion chains on adjacent rf minimum paths are shuttled into the same interaction zone. For coupling along the axial direction ($z$-direction), two ion chains on the same rf minimum path are brought close to each other. The above arrangement thus allows for transport of ions in axial direction while adjusting their radial distance to ions on other rf minimum paths. This ``radial transport" of ions is done by exclusively using dc voltages, removing the need for dynamic adjustments in the rf voltages in the QSA architecture scheme. The transition zones share some similarities with conventional rf junctions such as a change in the pseudo-potential confinement and axial potential barriers. However, contrary to conventional rf junctions~\cite{Burton2023}, the total confinement does not decrease inside the transition zones.\\

This paper is structured as follows: In section~\ref{sec:Design}, the trap concept and the design of the curved rf electrodes are presented. Then, in section~\ref{sec:Fabrication}, the fabrication of the demonstrator is described and the results of the characterization of the ion trap are presented. Finally, an outlook on future scaling of this type of ion trap chip is given in section~\ref{sec:Outlook}.

\section{Trap design}\label{sec:Design}

\begin{figure*}
  \captionsetup[subfloat]{position=top}
  \centering
  \subfloat[a][]{\includegraphics[scale=.75]{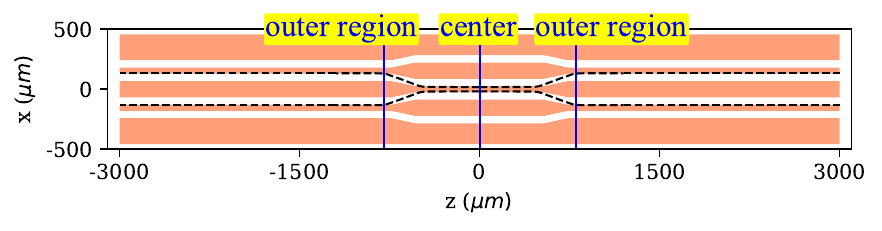} \label{fig:Lin_MT_full}} \\
    \subfloat[b][]{\includegraphics[scale=0.5]{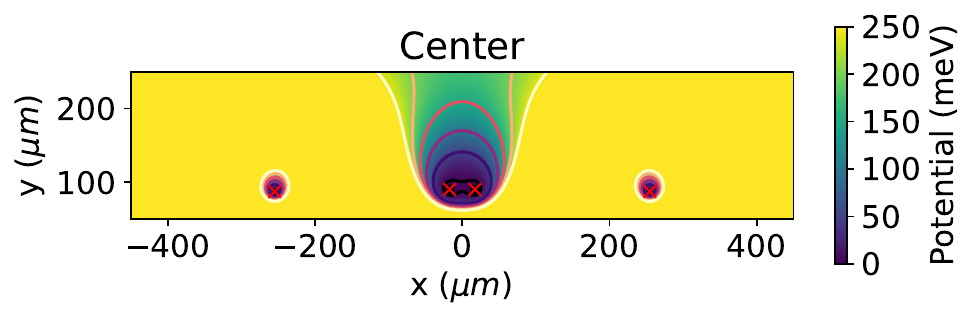} \label{fig:Lin_Meander_trap_pot_center}} 
    \hfill
  \subfloat[c][]{\includegraphics[scale=0.5]{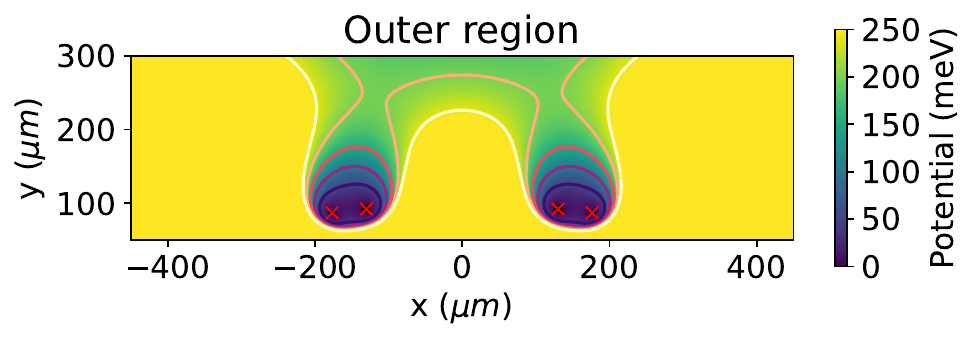}\label{fig:Lin_Meander_trap_pot_outer}}\\
  \subfloat[d][]{\includegraphics[scale=.75]{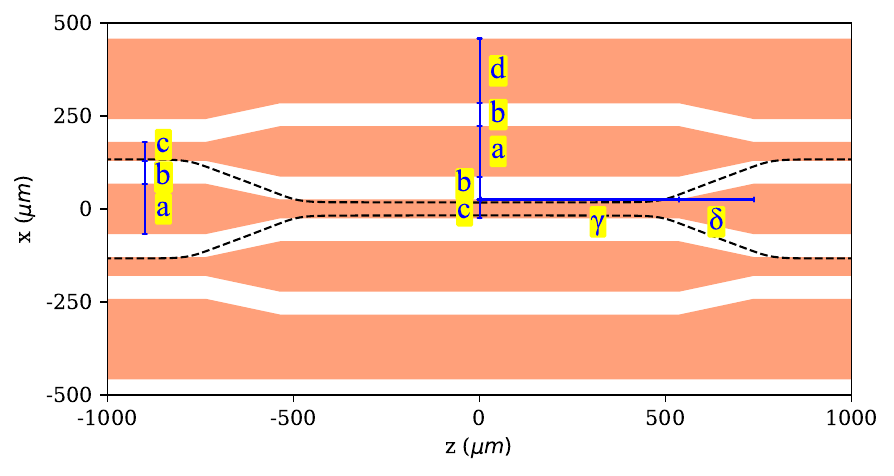}\label{fig:Lin_MT_center}}\\
  \subfloat[e][]{\includegraphics[scale=.75]{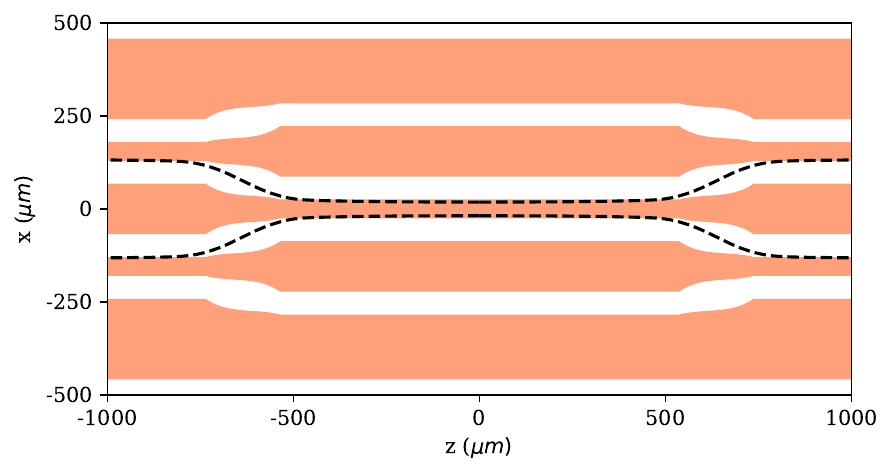}\label{fig:MT_center}}
\caption{Rf electrodes of the trap with linear and optimized transition zones and corresponding double-well potentials. The red areas are the rf electrodes, the white areas are spaces for dc electrodes. Rf minimum paths of the innermost double well are plotted as black dashed lines. a)~Full length of the rf electrodes with linear transition zones. The markings correspond to the potential cross-sections in the figures on the right. The line ``center" marks the center of the trap at $z = 0$. The lines ``outer region" mark the positions $z = \pm 800$~µm. Subfigures b, c are rf pseudo-potential cross-sections of the trap in the linear transition zone configuration in a) at different axial positions~$z$. The rf minima are marked by red crosses. b)~Rf pseudo-potential cross-section in the center of the trap at $z~=~0$~µm. This position is marked in a) by ``center". c)~Pseudo-potential in the ``outer regions" at $z~=\pm 800$~µm. This position is marked in a) by ``outer region". d)~Center of the trap with linear transition zones with markings for the different electrode dimensions. e)~Center of the trap with optimized transition zone shape.}\label{fig:Lin_MT} 
\end{figure*}

The ion trap chip consists of the two types of zones described above: interaction zones and transition zones. In the interaction zones, the rf electrodes are close to each other and linear, i.e., parallel to the rf minimum paths. The transition zones connect the interaction zones and have a shape optimized for minimal pseudo-potential barrier and change in total confinement, as described in subsection~\ref{sec:Transition_zone_optimization}. In the simplest case, shown in figure~\ref{fig:Lin_MT_full} and \ref{fig:Lin_MT_center}, the transition zones are just straight lines that connect the interaction zones. We will call this type of transition zones ``linear transition zones". 

In the interaction zones, two of the rf minima are close together, forming a double-well potential with minimal distance. This is illustrated in figure~\ref{fig:Lin_Meander_trap_pot_center}. The two minima should be, for calcium ions, at a distance of about $30 - 50$~µm. That allows, in accordance to equation~\ref{eq:coupling_rate}, double-well coupling in the kHz range. On the other hand, rf minima from other rf minimum paths should be kept sufficiently far away, typically in the hundreds of micrometers, so that undesired interactions with the former two minima are negligible. This is illustrated in figure~\ref{fig:Lin_Meander_trap_pot_outer}.

\subsection{Linear interaction zones}\label{sec:Lin_zones}

We started the design of the demonstrator with the linear parts (linear interaction zones) of the rf electrodes and by connecting these linear parts with straight lines (linear transition zones). The layout of this design is shown in figure~\ref{fig:Lin_MT}a. The optimized electrodes dimensions are shown in figure~\ref{fig:Lin_MT}b. We adopt a Cartesian coordinate system $(x, y, z)$, where $x, z$ lie in the plane of the surface of the trap, as shown in figure~\ref{fig:Lin_MT}, and $y$ is perpendicular to the surface.

In the center, shown in figure~\ref{fig:Lin_MT_center}, the layout consists of three rf electrodes that create a radial double-well potential, similar to the traps for radial coupling in \cite{Valentini2025}, and of two additional rf electrodes that increase the radial confinement and potential depth of this configuration. In the outer parts (in $z$-direction), the rf electrodes define two radial double-well potentials. The tunable design parameters of this configuration are the electrode widths $a$, $b$, $c$, $d$, the distance between the center of the trap and the transition zones $\gamma$ and the length of the transition zones $\delta$ (figure~\ref{fig:Lin_MT_center}). 

In the analysis, optimization and characterization of the trap designs, we used the gapless plane approximation \cite{Oliveira2000, House2008, Wesenberg2008}, implemented in the \texttt{electrode} package in Python \cite{Electrode-package}, based on \cite{SurfacePattern-package}. Dimensions stated in this part of the paper are therefore to be understood neglecting gaps between the rf and dc electrodes.

The axial rf field from the transition zones influences the ion position in the center. For the optimization of the electrode widths $a$, $b$, $c$ and $d$, we therefore fixed the transition zone parameters to $\gamma~\approx~500$~µm for the distance between the center of the trap and the transition zones and $\delta~\approx~200$~µm for the length of the transition zones. The choice for the value of $\gamma$ is the result of a trade-off between reducing the residual rf field from the transition zones in the center and, at the same time, making the distance between the center and the transition zones not too large for quick ion shuttling.

We chose as optimization constraints the positions of the minima $(\pm x_{0}^{\mathrm{c}}, y_{0}^{\mathrm{c}}, z = 0)$ of the central double-well potential. We minimized the Euclidean distances between the radial coordinates of the desired position of the minimum $(x_{0}^{\mathrm{goal}}, y_{0}^{\mathrm{goal}})$ and the actual position $(x_{0}^{\mathrm{c}}, y_{0}^{\mathrm{c}})$ for given rf electrode sizes as cost-function:
\begin{multline}
    \mathcal{C}(a, b, c, d) = 2\left(x_{0}^{\mathrm{c}}(a, b, c, d) - x_{0}^{\mathrm{goal}}\right)^{2}\\ + \left(y_{0}^{\mathrm{c}}(a, b, c, d) - y_{0}^{\mathrm{goal}}\right)^{2},
\end{multline}
with the optimization objectives $x_{0}^{\mathrm{goal}} = 18$~µm and $y_{0}^{\mathrm{goal}} = 90$~µm. The optimization objectives $x_{0}^{\mathrm{goal}}$, $y_{0}^{\mathrm{goal}}$ are similar in size to the minimum positions of the radial double-well in the trap for radial coupling in \cite{Valentini2025}. The value of $x_{0}^{\mathrm{goal}}$ corresponds to the actual resulting $x_{0}^{\mathrm{c}}$ when the transition zones are straight connection lines. In the case of curved transition zones, the value of $x_{0}^{\mathrm{c}}$ will be slightly modified by the different axial rf field. This slight modification is no fundamental issue.

The horizontal position of the minimum has a bigger weight compared to the ion-surface distance $y_{0}$, since the distance $s_{0} = 2x_{0}^{\mathrm{c}}$ between the minima of the radial double-well potential defines the strength with which two wells can couple along the radial direction according to equation~\ref{eq:coupling_rate} \cite{Valentini2025, Brown2011, Harlander2011}. We heuristically chose double the weight for $x_{0}$ than for $y_{0}$, since with this choice the optimization was more likely to converge to the desired value of $x_{0}^{\mathrm{goal}}$ than with equal weights. 

The ion-surface distance $y_{0}$ is relevant in terms of anomalous heating~\cite{Boldin2018}, but is not critical to the coupling rate~$\Omega_{\mathrm{ex}}$. We chose $y_{0}^{\mathrm{goal}} = 90$~µm to have similarly low heating rates as for the traps in reference~\cite{Valentini2025}, assuming a similar scaling of the heating rate in the ion-surface separation. 

Furthermore, we chose the distance between the rf electrodes $b \geq 60$~µm to obtain sufficient available space between the rf electrodes for the placement of dc electrodes. The optimization was then performed with a Nelder-Mead solver and non-negative random seed values. Since the optimization objectives $x_{0}^{\mathrm{goal}}$, $y_{0}^{\mathrm{goal}}$ are two values that do not constrain the electrode sizes $a$, $b$, $c$, $d$ completely, we compared the solutions obtained from different seed values in terms of pseudo-potential depth in the center and rf electrode sizes. We chose the trap dimensions leading to the highest pseudo-potential depth and largest electrode sizes in the set. From this procedure, we obtained values of $a~=~135.6$~µm, $b~=~61.5$~µm, $c~=50.8$~µm and $d~=174$~µm. This results for ${}^{40}\mathrm{Ca}^{+}$ in a pseudo-potential depth of 178.4~meV and a potential barrier between the wells in the center of 5.1~meV at 194.5~V rf amplitude and an rf drive frequency of $2\pi\times 31.91$~MHz. These values of the rf voltage and frequency are the ones that were experimentally used in the characterization of the trap reported in section~\ref{sec:Fabrication}. 

The corresponding rf pseudo-potential at the center of the trap and outside the transition zones is shown in figures~\ref{fig:Lin_Meander_trap_pot_center} and \ref{fig:Lin_Meander_trap_pot_outer}.
In figure~\ref{fig:Lin_Meander_trap_pot_center}, one can see the pseudo-potential in the center of the trap at $z = 0$. The two inner minima are $36$~µm apart in a double-well. At this separation, the coupling is $\Omega_{\mathrm{ex}} = 2\pi\times 5.9$~kHz at a mode frequency of 1~MHz, while the two outer minima are isolated. In figure~\ref{fig:Lin_Meander_trap_pot_outer} on the other hand, the pseudo-potential in the outer parts of the trap, at $z = 800$~µm, is shown. Here, the inner minima are apart and form a double-well with increased coupling rate with the outer minimum on each side.

\subsection{Transition zone optimization}\label{sec:Transition_zone_optimization}
\begin{figure*}[t]
    \centering
    \subfloat[a][]{\includegraphics[scale=0.5]{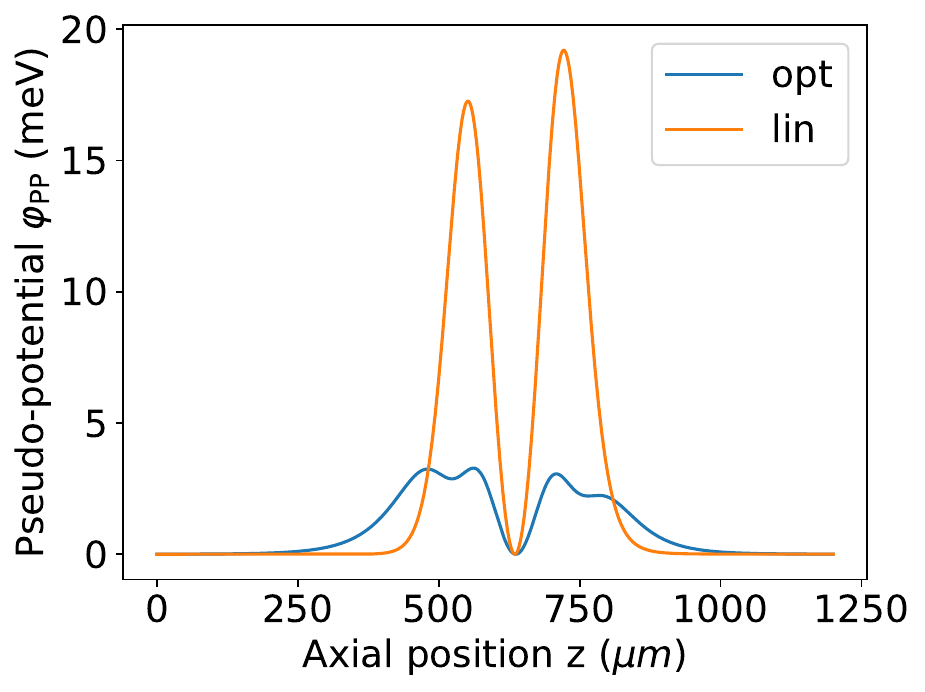}\label{fig:PP}}
    \subfloat[b][]{\hspace{0.45cm}\includegraphics[scale=0.5]{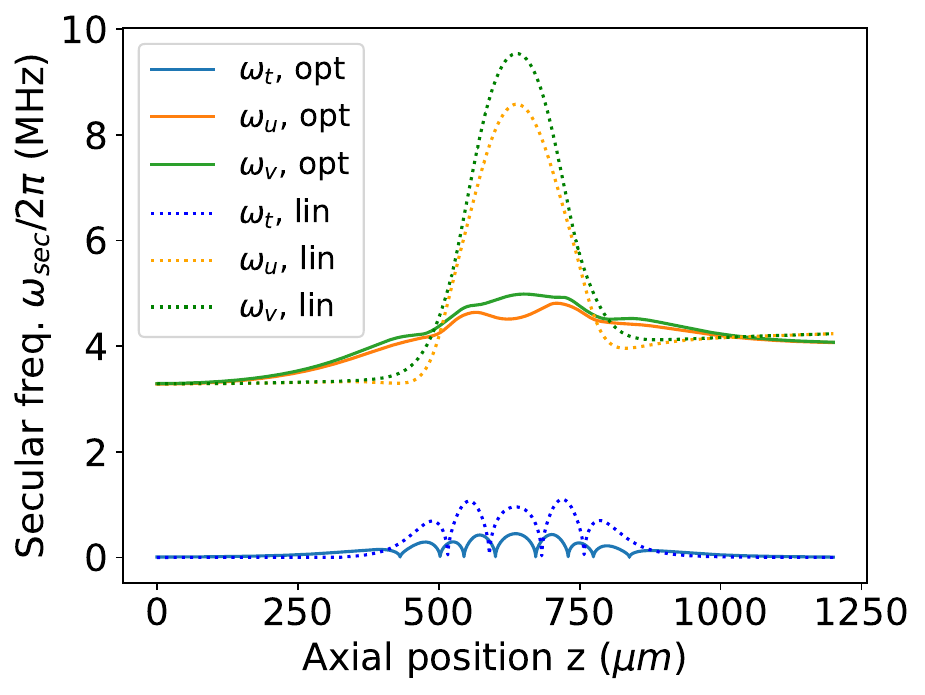}\label{fig:sec_freq}} \\
    \subfloat[c][]{\includegraphics[scale=0.5]{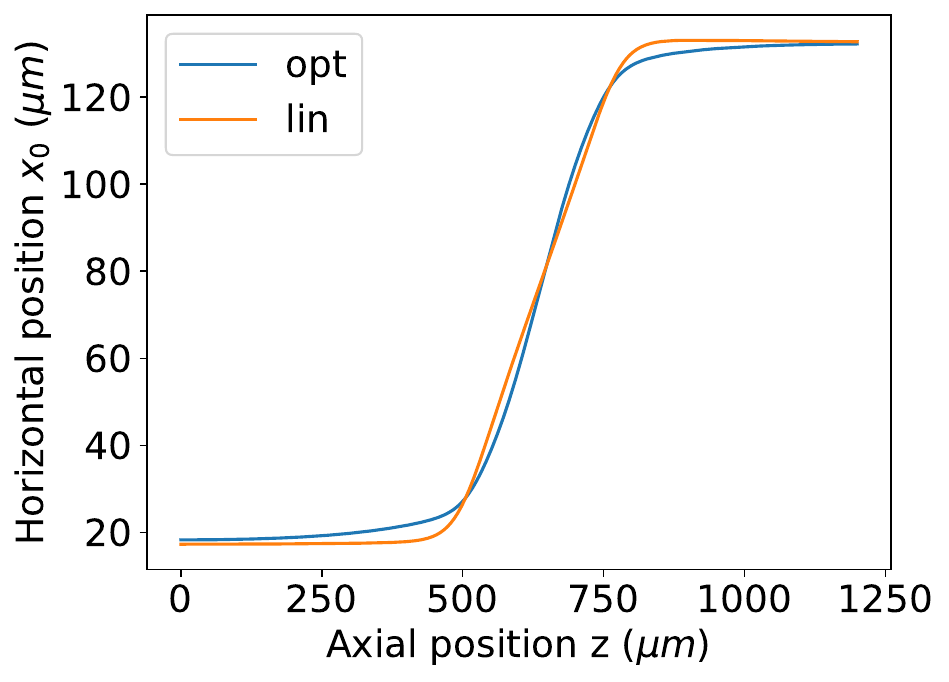}\label{fig:horiz_pos}}
    \subfloat[d][]{\includegraphics[scale=0.5]{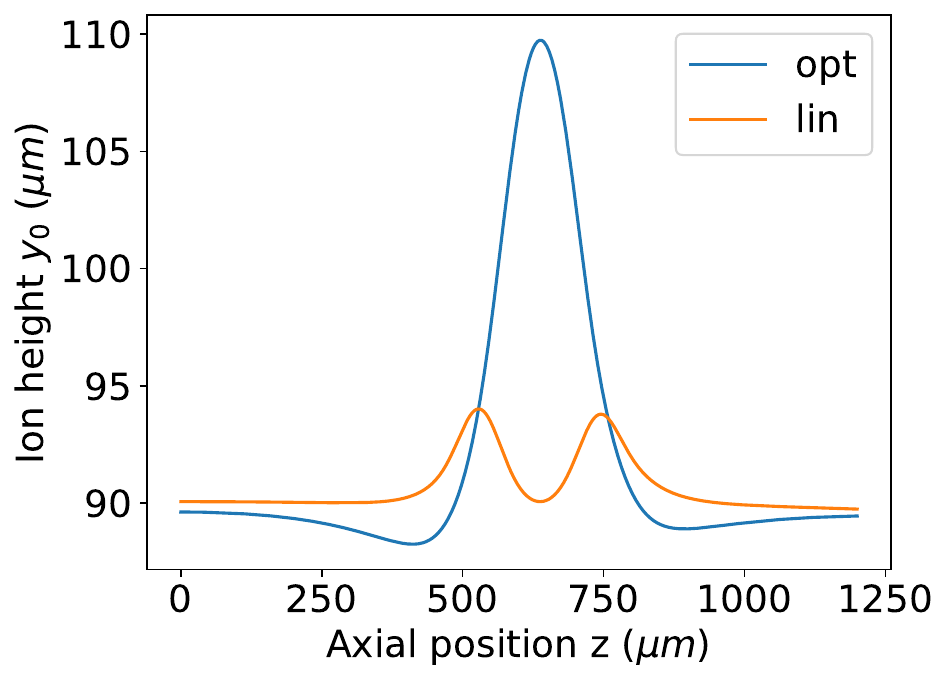}\label{fig:height}}
    \caption{Pseudo-potential $\varphi_{\mathrm{PP}}$, secular frequencies $\omega_{i}$, horizontal minimum position $x_{0}$ and ion-surface distance $y_{0}$ as a function of axial position $z$ for linear and optimized transition zones. The values are along the rf minimum path $\vec{\xi}(z) = (x_{0}(z), y_{0}(z), z)$. All calculations assume 194.5~V rf amplitude and an rf drive frequency of $2\pi\times 31.91$~MHz. a) Tangential pseudo-potential along the rf minimum path. b) Secular frequencies of the eigenmodes, considering only rf electrode contributions. c) Horizontal position of the minimum $x_{0}$, which is half the ion-ion distance $s_{0}$ in the central double-well. d) The ion-surface separation $y_{0}$ as function of the axial position~$z$.}
\end{figure*}
After optimizing the linear parts of the trap, we optimize the transition zones between them. We start by studying the transport through the linear transition zones, to determine if a more sophisticated shape is beneficial. Only rf confinement and no dc potentials are considered.

We fixed the transition zone parameters to be $\gamma~\approx~500$~µm for the distance between the trap center and the transition zones and $\delta~\approx~200$~µm for the length of the transition zones. In the transition zones, one expects the largest residual axial rf field -- \textit{axial} meaning here along the rf minimum path -- because the rf electrodes are curved there and induce unwanted micromotion~\cite{Zhang2022, Blakestad2011}. Therefore, one wants to make the transition zones as short as possible, so that an ion, at a given velocity, can traverse it as quickly as possible. Moreover, enough space is needed inside the transition zones to place dc electrodes for sufficient control of the ions during shuttling. We opted for $\delta = 200$~µm to allocate enough space for five independent dc electrodes of $40$~µm length, while keeping the length of the transition zone small to minimize ion heating.

There are several ways that one can choose to transport an ion through a region where the rf electrodes change their shape, such as transport along the minimum of the rf potential or transport along the path of constant total confinement~\cite{Burton2023, Taniguchi2025}, where the sum of the squares $\omega^{2}~=~\sum_{i}\omega_{i}^{2} = (1/m)\vec{\nabla}^{2}\varphi_{\mathrm{PP}}$ of the secular frequencies $\omega_{i}$ ($i = x, y, z$) is kept constant. The total confinement $\omega^{2}$ depends solely on the rf fields, since the dc fields only distribute the confinement along the secular modes, i.e., they have zero divergence~\cite{Burton2023, Zhang2022}.

For optimizing the shape of the transition zones, we chose to consider transport along rf minimum paths, which are the paths of minimal pseudo-potential. Transport along the path of minimal pseudo-potential is discussed in \cite{Zhang2022} and \cite{Burton2023, Taniguchi2025} for the case of X-junctions, with the goal of minimizing axial micromotion along the transport path~\cite{Burton2023}. 

The position of an ion in an X-junction can be satisfactorily described by its position in the lab frame of reference $(x, y, z)$. In the transition zones considered here, the rf minimum path between the center and the outer linear parts is curved. We therefore consider the ion transport in a frame of reference $(t, u, v)$, moving along the rf minimum path $\vec{\xi}(z) = (x_{0}(z), z, y_{0}(z))$, parametrized by the axial position $z$. Here, $x_{0}(z), y_{0}(z)$ are the other coordinates of the rf minimum on this rf minimum path for a given $z$. The direction $\hat{t} = \dot{\xi}/|\dot{\xi}|$ is chosen to be the tangent to the rf minimum path $\vec{\xi}$, while the directions $\hat{u}$, $\hat{v}$, normal to $\hat{t}$, are chosen to be the radial eigenmodes of oscillation, i.e., the normalized eigenvectors of the Hessian $(H\varphi_{\mathrm{PP}})_{ij} = \partial_{i}\partial_{j}\varphi_{\mathrm{PP}}$.

The pseudo-potential along the rf minimum path $\varphi_{\mathrm{PP}}(\vec{\xi}(z))$ can be interpreted as the local ``axial" pseudo-potential barrier. 
We find that the maximum pseudo-potential barrier for the linear transition zone configuration, shown in figure~\ref{fig:PP}, is 18 meV. This value can exceed barrier heights of radial double-wells that are typically at 1 - 5 meV. If the barrier height is lower than the pseudo-potential at the minima of the double-well, it can lead to ion hopping between the wells.

For coherent ion transport, not only the change in the pseudo-potential is relevant, but also the change in confinement and secular frequencies. The radial secular frequencies $\omega_{u}$, $\omega_{v}$, plotted in figure~\ref{fig:sec_freq}, can be more than three times greater inside the linear transition zones than outside. Note that the radial secular frequencies in the linear parts of the trap are nearly degenerate since, at this point, only the rf contributions, without dc electrodes to lift the degeneracy, are considered. Inside the transition zones, the degeneracy is lifted by the rf electrode geometry. The total confinement $\omega^{2}(z) = \omega_{t}^{2} + \omega_{u}^{2} + \omega_{v}^{2} = \mathrm{tr}(H\varphi_{\mathrm{PP}})/m = \vec{\nabla}^{2}\varphi_{\mathrm{PP}}(x_{0}(z), z, y_{0}(z))/m$ is also larger inside the linear transition zones than outside, because of the larger secular frequencies. In contrast to the similar optimization of X-junctions, the issue of steeply decreasing total confinement \cite{Zhang2022, Wesenberg2009, Burton2023, Wright2013, Blakestad2009, Taniguchi2025} does not arise. Since the total confinement at a point is determined by the rf electrode geometry, rf voltage and rf frequency alone \cite{Zhang2022} and cannot be changed by the dc electrodes \cite{Burton2023}, it is desirable to keep it as constant as possible in the transition zone. In contrast to X-junctions, for the transition zones discussed here, the total confinement \textit{increases} inside transition zones. For the transition zones discussed here, keeping the total confinement as constant as possible thus means constraining its increase.

From the above considerations, we find that in an optimal geometry the axial pseudo-potential barrier should be as low as possible, while the total confinement should stay as constant as possible or at least not increase excessively. The optimal transition zone geometry is given by the shape of the rf electrodes. To fix notation, we describe the boundary of a transition zone by a cubic B-spline that is defined by a list of its control points $\mathbf{B}$.

We optimized the shape of the transition zones with a cost function $\mathcal{C}$ consisting of two terms:
\begin{equation}
    \mathcal{C}\left(\mathbf{B}\right) = w_{1}\int_{\gamma}^{\gamma+\delta}\omega^{2}(\vec{\xi}(z))\mathrm{d}z + w_{2}\int_{0}^{z_{\mathrm{F}}}\left\vert\frac{\partial\varphi_{\mathrm{PP}}}{\partial\hat{t}}\right\vert\mathrm{d}z,\label{eq:cost_transition_zone}
\end{equation}
where $\omega^{2}(\vec{\xi}(z)) = \omega^{2}(x_{0}(z), y_{0}(z), z)$, $\partial\varphi_{\mathrm{PP}}(z) / \partial\hat{t} = \hat{t}(z)\cdot\vec{\nabla}\varphi_{\mathrm{PP}}\left(x_{0}(z), y_{0}(z), z\right)$, $z_{\mathrm{F}} = \gamma + 500$~µm, and $w_{1}$, $w_{2}$ are weights.

The first term in equation~\ref{eq:cost_transition_zone} constrains the magnitude of the total confinement in the transition zone ($z=\gamma$ to $z=\gamma+\delta$). For most transition zone shapes, the confinement $\omega^{2}$ tends to increase rapidly inside it. Constraining $\omega^{2}$ in the transition zones then keeps it as constant as possible. 
 
The second term in equation~\ref{eq:cost_transition_zone} is the same as that used in \cite{Zhang2022, Mokhberi2017}. It suppresses the tangential pseudo-potential gradient \cite{Mokhberi2017} that can cause ion heating during shuttling~\cite{Sedlacek2018, Blakestad2009}. By minimizing this term, the pseudo-potential barrier is reduced~\cite{Zhang2022}. Since the pseudo-potential barrier and pseudo-potential gradient are desired to be small everywhere along the rf minimum path, the second term spans not only the transition zone, but the whole region of interest from $z = 0$ to $z_{\mathrm{F}} = \gamma + 500$~µm, where the value of 500~µm is an educated guess to make sure also the pseudo-potential in the outer parts of the trap is covered.

The shape of the transition zones was described by a cubic B-spline \cite{Zhang2022, Mokhberi2017} with, at first, three internal control points, between which the shape is interpolated by the spline, and two end points at $(\gamma, c/2)$ and $(\gamma+\delta, a/2)$. We also tried different numbers of internal control points and parameterizing the boundary of the transition zones by many ($\sim 20 - 100$) single connected points. In the end, two internal control points of a cubic B-spline turned out to be the most computationally efficient description of a smooth transition zone boundary. Furthermore, to define both boundaries of the transition zone, we imposed an 180°-inversion symmetry around the center of the transition zone such that the shape of the transition zone looks the same for an ion going from the center of the trap to the outer part and vice versa. 

The optimization was performed with a Nelder-Mead solver. The terms in the cost function, equation~\ref{eq:cost_transition_zone}, were evaluated at 5000 evenly spaced points between $z = 0$ and $z = z_{\mathrm{F}}$ on the rf minimum path. The control points were initialized with 3 points that were evenly spaced on the original linear transition zone boundary and, as a precaution, shifted randomly by $\pm 15\%$ in the $x$- and $z$-directions to avoid getting trapped in a single local minimum. 
We picked the local minimum with the lowest value of the cost function that simultaneously fulfilled the fabrication and design constraints. The constraints are that the rf electrodes do not get too narrow to keep the resistance low (larger than 20~µm) and leave enough space for placing dc electrodes ($b(z) > 30$~µm for $\gamma < z < \gamma+\delta$) in the transition zone.

The layout of the rf electrodes of the trap, resulting from the above optimization with equal weights $w_{1} = w_{2}$, is shown in figure~\ref{fig:MT_center}. The corresponding optimized pseudo-potential is shown in figure~\ref{fig:PP}. One sees that the resulting pseudo-potential barrier is less than a fifth of the barrier of a linear transition zone. Similarly, the increase of the radial secular frequencies is only a quarter that of the linear case (figure~\ref{fig:sec_freq}). The horizontal ion position $x_{0}$ and with it the ion-ion separation $s_{0}=2x_{0}$ along the $z$-axis are modified by 1~µm by the non-linear shape of the transition zone (figure~\ref{fig:horiz_pos}). The ion-surface separation $y_{0}(z)$ is now increased in the transition zone compared to the linear case (figure~\ref{fig:height}). This increase in the ion-surface separation, however, did not pose a challenge to the experimental characterization of the trap. The ion-ion separation in the center is slightly modified to $s_{0} = 36.6$~µm and the ion-surface separation to $y_{0} = 89.6$~µm.

\subsection{Electrode routing\label{electrode_routing}}
\begin{figure}
    \centering
    \includegraphics[width=\linewidth]{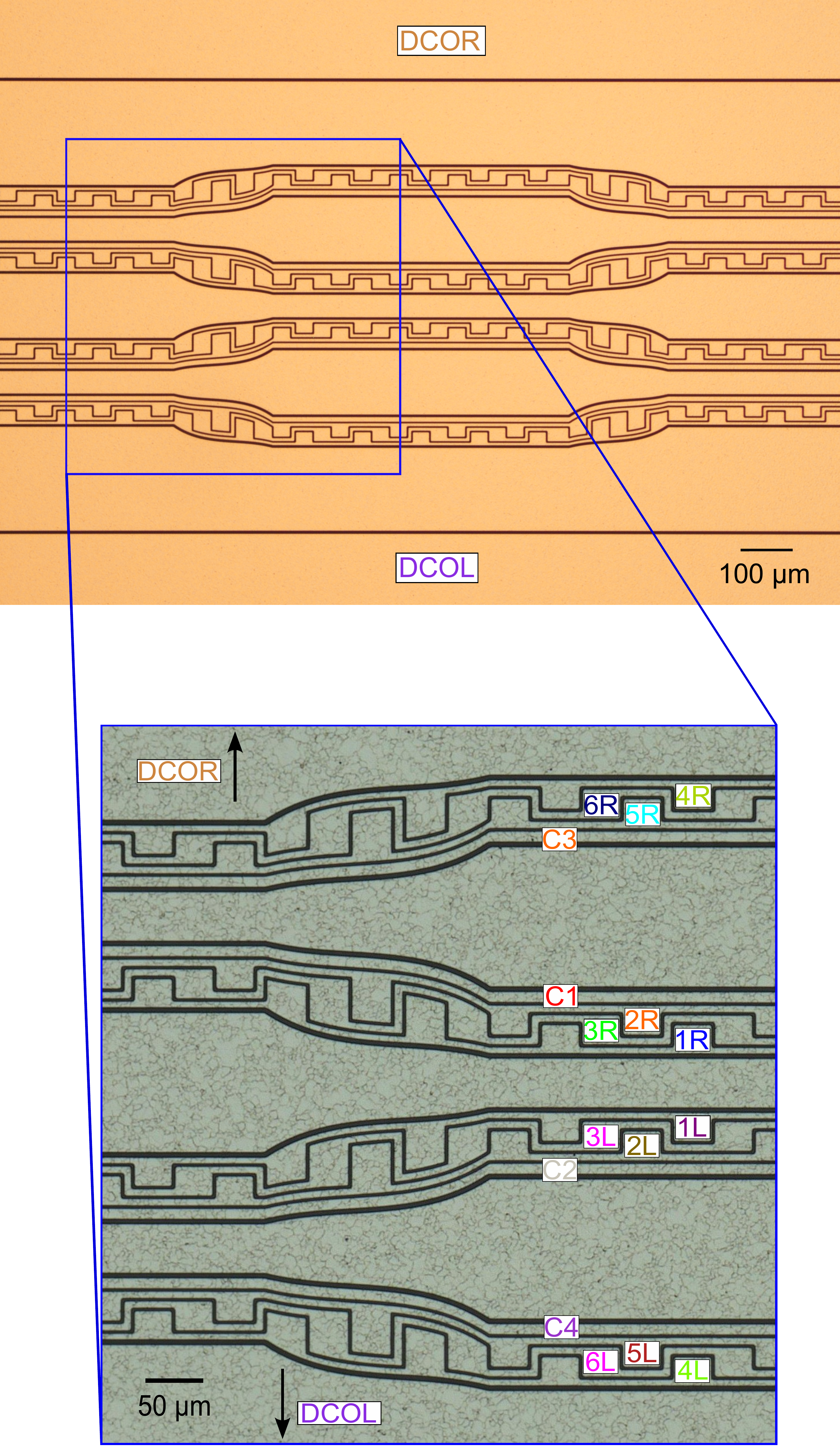}
    \caption{Optical microscopy image of the center of the ion trap chip. One can see the ``meander-shaped" dc electrodes. Blue box: Optical microscopy image of the ``meander-shaped" dc electrodes with labels between the rf electrodes in the interaction and the transition zones. Indicated by arrows, but not shown in the blue box, is one rectangular dc electrode outside of the rf rails on each side (DCOR, DCOL).}
    \label{fig:MT_dcs}
\end{figure}

We opted for fabrication of the ion trap chip as a single-metal-layer trap to allow for rapid fabrication, which constrains the possible dc electrode shapes and connections that can be routed for control of ions. 

For the dc electrodes, which are used to create axial/tangential confinement, we chose to connect the dc electrodes periodically using a saw-tooth like (``meander") pattern. An image of the center of the single-metal-layer ion trap chip can be seen in figure~\ref{fig:MT_dcs}. The dc electrodes with the ``meander" pattern~\cite{Holz2020-2, Anmasser2025} can be seen in the blue box in figure~\ref{fig:MT_dcs}. This routing follows the same principle as the ion trap chip in \cite{Holz2020}.
This ``meander" pattern allows three dc electrodes to act as a periodic arrangement of dc electrodes~\cite{Holz2020-2}. Additionally, a long compensation electrode (C1-4) in each rail allows for stray field compensation in $x$-direction. Furthermore, outside the rf electrodes is one large dc electrode (DCOR/L) on each side of the trap for efficient stray field compensation in the vertical $y$-direction.

For 1~MHz of axial confinement, the largest dc voltage has about 12~V magnitude. The dc voltages for stray field compensation, for 1~V/mm of stray field, are all in the range of $\pm 2.5$~V. 

\section{Trap fabrication and characterization}\label{sec:Fabrication}
Ion trap chips with the electrode layout presented in section \ref{sec:Design} have been fabricated at Infineon Technologies in Villach, Austria, using industrial microfabrication processes. The chips were fabricated in the same facilities as the chips reported in \cite{Holz2020, Auchter2022, Dietl2025, Anmasser2025} and \cite{Valentini2025}. The chips were fabricated on 200~mm wafers using CMOS-based fabrication process flows~\cite{Holz2020, Dietl2025, Auchter2022}.

Fused silica wafers with polycrystalline silicon on the backside were used as a substrate to facilitate processing in tools optimized for silicon wafers~\cite{Dietl2025}. Fused silica offers an interesting platform for prototyping of ion trap electrode designs. Compared to a silicon substrate, no shielding layer between the rf electrodes and the substrate is needed, because fused silica shows low rf dissipation~\cite{Dietl2025, Sterk2024} and has a band-gap~\cite{Dietl2025} larger than the photon energy of lasers used in typical ${}^{40}\mathrm{Ca}^{+}$ experiments~\cite{Roos2000}. Since there is no need for a shielding layer, the capacitance between the rf electrodes and ground is lowered, leading to lower power dissipation in the rf electrodes~\cite{Dietl2025}.

The electrodes of the trap were fabricated by depositing a single layer of aluminum of 2~µm thickness on fused silica using sputter deposition. This layer was structured by I-line photolithography~\cite{Greeneich1988}, utilizing stepper reticles and plasma etching~\cite{Dietl2025}. 
More details on the fabrication processes and steps involved can be found in references~\cite{Dietl2025} and \cite{Holz2020}. 

We now present the characterization of the single-layer ion trap chip with the rf electrode design described in section~\ref{sec:Design}. The ion trap was characterized by determining the stray electric field and the micromotion modulation index in the center. For the characterization of the core functionality of the trap, we transported ions from the center through the transition zones and back. 

The experiments were conducted in a cryogenic ion trapping setup at a base temperature of about 7~K~\cite{Anmasser2025}. The trap was operated at around 195~V rf amplitude and an rf drive frequency of 32~MHz. The experiments were performed with ${}^{40}\mathrm{Ca}^{+}$ ions, with the $4\mathrm{S}_{1/2} \leftrightarrow 3\mathrm{D}_{5/2}$ quadrupole transition for thermometry, driven by a 729~nm laser and the $4\mathrm{S}_{1/2} \leftrightarrow 4\mathrm{P}_{1/2}$ dipole transition at 397~nm for Doppler cooling and detection~\cite{Roos2000, Leibfried2003}. The calcium is supplied by evaporation from a calcium oven. A Fastino DAC card~\cite{Sinara2025} was used for supplying the dc voltages to the experiment.

\begin{figure}
  \centering
  \subfloat[a][]{\includegraphics[scale=0.45]{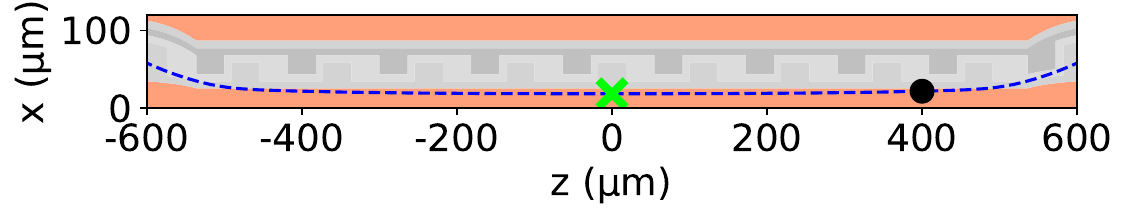}\label{fig:MT_min_lin}} \\
  \subfloat[b][]{\includegraphics[scale=0.45]{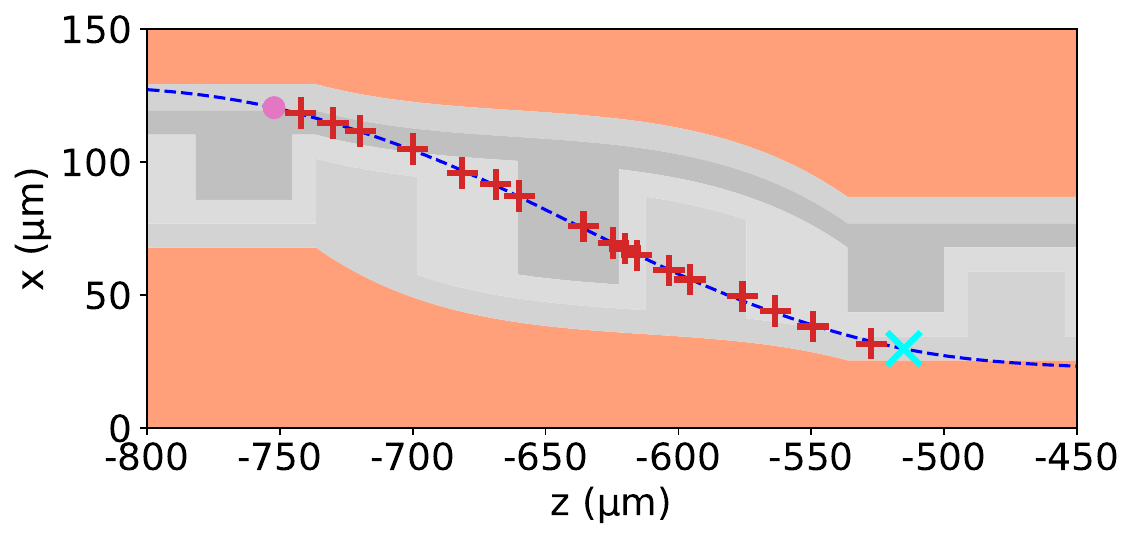}\label{fig:MT_min_trans}}
  \caption{Horizontal projections of the shuttling along the rf minimum path on the trap surface. a) Rf minimum path (blue dashed line) with starting point (green cross) and end point (black circle). b) Horizontal projection of the rf minimum path through a transition zone on the trap surface (blue dashed line). The measured shuttling process started at the cyan cross and ended at the pink circle. Red points in between are measured ion positions. The points are plotted without error bars for better visibility. The uncertainty of each point, given by the CCD camera read-off error, is $\pm 2.3$~µm in each direction.} \label{fig:MT_char_min}
\end{figure}

First, we verified that we are able to reliably trap ions in the center of the trap and along the two inner rf minimum paths on both sides of the trap. We then measured heating rates in the trap. Heating rates in the upper center trapping site at $(x_{0}, z_{0}, y_{0}) = (18.3, 0, 89.6)$~µm (see green cross in figure~\ref{fig:MT_min_lin}) were measured using sideband thermometry~\cite{Leibfried2003}. We measured a heating rate of $18 \pm 3$~ph/s at an axial secular frequency of 1.2~MHz. 

The central functionality of the presented ion trap chip is the ability to shuttle from one interaction zone to the next. To investigate as a first step the dc shuttling in the central linear region of the trap, we first shuttled an ion from the upper center ($x_{0}~=~18.3$~µm) at $z~=~0$~µm to $z~=~400$~µm and back. This shuttling path is illustrated in figure~\ref{fig:MT_min_lin} by the blue dashed line. The outer point is at an axial position before the rf minimum paths start to bend toward the transition zones. We shuttled the ion in 15 steps in each direction with 800~µs wait time per step, which results in a total shuttling time of 24~ms. With a total shuttling distance of 0.8~mm, this corresponds to a shuttling velocity of 3.3~cm/s. 
The shuttling velocity in our experiment was limited by RC filters (cutoff-frequency 1~kHz) on the dc filter board. We started each shuttling attempt manually and checked that the ion has indeed returned on the camera image. The ion returned from $z~=~400$~µm to $z~=~0$~µm in 80 of 80 shuttling attempts. This results, using the rule of succession~\cite{Ross2020}, in a lower bound for the return probability of 98.8~\%.

Next, to prove the full functionality of the proposed shuttling mechanism, we shuttled an ion from the upper center at $z~=~0$~µm, $x_{0}~=~18.3$~µm (green cross in figure~\ref{fig:MT_min_lin}) to the upper outer part at $z = -800$~µm, $x_{0} = 127.3$~µm (pink circle in figure~\ref{fig:MT_min_trans}) and back. That is, shuttling the ion from the trap center through a transition zone and back. The shuttling path through the transition zone can be seen in figure~\ref{fig:MT_min_trans}. We shuttled the ion in 200 steps in each direction with 4~ms wait time per step, leading to a total shuttling time of 800~ms. The total shuttling distance is 1.6~mm, resulting in a shuttling velocity of 2~mm/s (measured along the $z$-axis). 
We observed occasional ion loss for faster shuttling speeds. The ion loss during shuttling could be caused by imperfections in the DAC unit or its control.
We started each shuttling attempt with a shuttling speed of 2~mm/s manually and checked that the ion has indeed returned on the camera image. The ion returned from the upper part behind the transition zone at $z = -800$~µm with the stated shuttling parameters in 70 of 70 cases, resulting in a lower bound for the return probability of 98.6~\%.

To characterize the ion path during shuttling through the transition zones, we trapped an ion at $z_{0} = -515$~µm, $x_{0} = 29.7$~µm and took pictures of its position with a CCD camera while adiabatically shuttling it to $z_{0} = -757$~µm, $x_{0} = 121.5$~µm, optimizing the micromotion compensation at each step. The rf minimum path between these points is shown in figure~\ref{fig:MT_min_trans} between the cyan and the pink circle. We moved the same ion in steps of 1~µm, adjusting the dc stray field compensation voltages and position of 397~nm detection and cooling lasers wherever necessary. We compared the position of the ions on the camera images with an image of the trap surface underneath. Ion positions recorded this way are marked in figure~\ref{fig:MT_min_trans} by red crosses. One can see that the ion position closely follows the rf minimum path, as expected and desired in this scenario.

Residual axial pseudo-potential in the trap results in increased micromotion. The residual axial potential is introduced in the trap center by the transition zones. The micromotion modulates the effective laser-ion interaction with the 729~nm thermometry laser. This modulation is measured by the micromotion modulation index~$\beta$~\cite{Berkeland1998, Keller2015}. The micromotion modulation index is a direct measure of the micromotion along the direction of the laser~\cite{Keller2015}. To quantify the influence of the residual axial pseudo-potential, we measured the micromotion modulation index~\cite{Keller2015} from the upper center of the trap at $z = 0$~µm to $z = 70$~µm in 10~µm steps. For this measurement, we first trapped an ion in the center ($z = 0$~µm) and optimized the compensation of its micromotion \cite{Berkeland1998} by minimizing the micromotion sideband of the 729~nm carrier~\cite{Roos2000, Pruttivarasin2014}. The compensated stray field in the center had the magnitude $\vec{E}_{\mathrm{stray}} = (E_{x}, E_{z}, E_{y}) = (-0.16, 0.46, 1.2)$~V/mm. The magnitude of the stray field~$\vec{E}_{\mathrm{stray}}$ was found by inserting the experimentally found stray field compensation dc voltages into the electrostatic simulation of the trap and calculating the corresponding electric field. 

\begin{figure}
    \centering
    \includegraphics[width=1\linewidth]{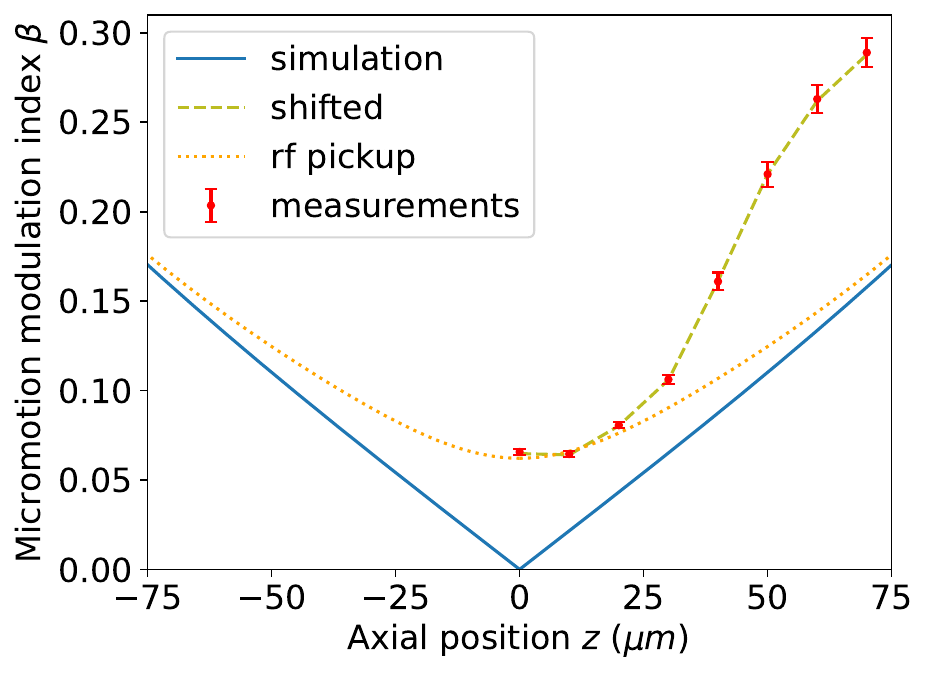}
    \caption{Micromotion modulation index $\beta$ measured in the upper center part ($x_{0} = 18.3$~µm) of the trap at $z = 0$ to $z = 70$~µm. In red are the values measured using ratios of Rabi frequencies and in blue is the micromotion modulation index expected from simulations with the ion on the rf minimum path without any stray fields. The orange dotted line is the $\beta$ expected from an rf pickup of 0.2~V on the dc electrodes 1R and 1L (cf. figure~\ref{fig:MT_dcs}). The green dashed line is the $\beta$ if one assumes the ion not on the rf minimum path, but shifted by a value $\alpha(z)$ in $x$-direction (radial) from it. The shift $\alpha$ increases with $z$, e.g. $\alpha (z = 0 ) = -86$~nm and $\alpha (z = 70~\mu m) = -320$~nm. See figure~\ref{fig:MM_shifts} for a plot of $\alpha (z)$.}
    \label{fig:MM}
\end{figure}

\begin{figure}
    \centering
    \includegraphics[width=\linewidth]{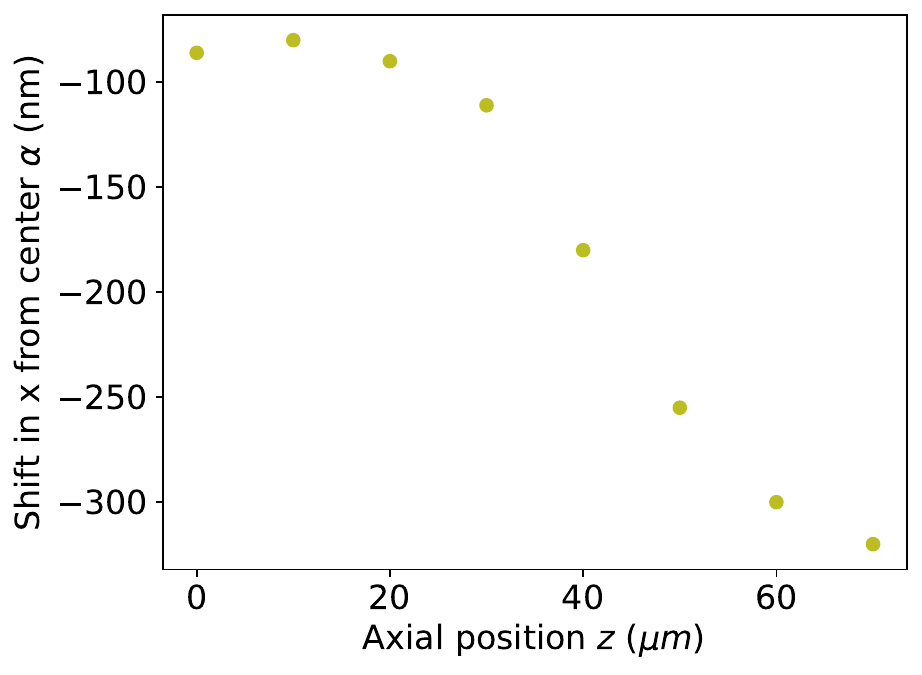}
    \caption{Calculated shifts~$\alpha$ as a function of the axial position~$z$- The shifts~$\alpha$ were calculated from the measured values of the micromotion modulation index~$\beta$ in figure~\ref{fig:MM}. The shift $\alpha$ increases with $z$, e.g. $\alpha (z = 0 ) = -86$~nm and $\alpha (z = 70~\mu m) = -320$~nm.}
    \label{fig:MM_shifts}
\end{figure}

Keeping the stray field compensation constant to see how the micromotion modulation index would vary along a long ion chain, we performed Rabi flops on the qubit transition and the corresponding micromotion sideband at the axial positions $z = 0...70$~µm. 
The micromotion modulation index $\beta$ at the different positions was determined using the Rabi frequencies on the carrier $\Omega_{\mathrm{carr}}$ and the micromotion sideband $\Omega_{\mathrm{mm}}$ by~\cite{Kiesenhofer2024, Keller2015}:
\begin{equation}
    \frac{\beta}{2} \approx \frac{J_{1}(\beta)}{J_{0}(\beta)} = \frac{\Omega_{\mathrm{mm}}}{\Omega_{\mathrm{carr}}},
\end{equation}
where $J_{1}, J_{0}$ are Bessel functions of the first kind. The values of $\beta$ found in this measurement are displayed as red points in figure~\ref{fig:MM}.

The blue solid, orange dotted and green dashed lines in figure~\ref{fig:MM} are predictions for the micromotion modulation index from electrostatic simulations. They are calculated from the pseudo-potential $\varphi_{\mathrm{PP}}$ (cf. figure~\ref{fig:PP}) by~\cite{Zhang2022, Josteinsson2023, Keller2015}:
\begin{equation}
    \beta(x_{0} + \alpha, z, y_{0}) = \frac{4\pi}{\Omega_{\mathrm{rf}}\lambda_{\mathrm{L}}}\sqrt{\frac{e\varphi_{\mathrm{PP}}(x_{0}(z) + \alpha, z, y_{0}(z))}{m}}\cos{\vartheta},\label{eq:mm}
\end{equation}
where $\alpha$ is an assumed shift from the rf minimum in $x$-direction (radial), $m$ the ion mass, $e$ the charge of a ${}^{40}\mathrm{Ca}^{+}$-ion, $\Omega_{\mathrm{rf}} = 2\pi\times 31.91$~MHz the rf drive frequency, $\lambda_{\mathrm{L}} = 729$~nm the wavelength of the laser driving the qubit transition and $\vartheta = \pi/4$ the angle between the axial ($z$) direction and the direction of the laser. 

One observes that the measured values of $\beta$ are larger compared to the results from the calculation for the ion on the rf minimum path without any stray fields (blue line in figure~\ref{fig:MM}), but also show an increasing trend with increasing axial position $z$. The residual axial rf field should be exactly zero in the trap center at $z = 0$ due to symmetry. The measured non-zero micromotion modulation index at the rf null at $z = 0$ thus indicates that the ion had some residual micromotion, either coming from rf pickup on the dc electrodes~\cite{Holz2020-2} or the ion not being exactly in the trap center, but a bit shifted from it. 

The orange dotted line in figure~\ref{fig:MM} is the micromotion modulation index~$\beta$ expected from rf pickup on the dc electrodes. For the calculation of the orange curve, it was assumed that the electrodes 1R and 1L (cf. figure~\ref{fig:MT_dcs}) each pick up 0.2~V of rf voltage from the rf electrodes next to them. The rf pickup value of 0.2~V was picked to explain the value of $\beta \neq 0$ at $z = 0$. Nevertheless, at $z > 0$, the measured values of $\beta$ are larger than the ones expected from an rf pickup of 0.2~V. 

A second hypothesis, besides rf pickup, is that the ion was not exactly on the rf minimum path due to stray electric fields. The stray fields would cause the ion to be shifted by $\alpha (z)$ from the rf minimum path. Since the rf fields in the $z$-direction are negligibly small for small $z \approx 0$, it is more likely that the residual fields stem from a shift in the $x$-direction, in which the pseudo-potential rapidly increases also in the center. The green dashed line in figure~\ref{fig:MM} is a fit of with a displacement function $\alpha (z)$ to the measured values of $\beta$, given by the calculated displacements from the rf minimum in $x$-direction. Inserting the measured value of $\beta$ at $z = 0$ into equation~\ref{eq:mm}, results in $\alpha = -86$~nm. For increasing $z$, the absolute value of the resulting $\alpha$ increases to $\alpha = -320$~nm at $z = 70$~µm. 
The green dashed line in figure~\ref{fig:MM} is a fit of $\alpha (z)$ to the data points for each $z$. The shift values $\alpha (z)$, fitted to the measured micromotion modulation indices $\beta (z)$, $z = 0...70$~µm, are plotted in figure~\ref{fig:MM_shifts} as a function of the axial position~$z$. 
From the presented measurement data, we can conclude that we were unable to completely eliminate the micromotion in the  center in the experiment.

\section{Multi-layer trap}\label{sec:Outlook}
\begin{figure*}
    \centering
    \includegraphics[width=1\linewidth]{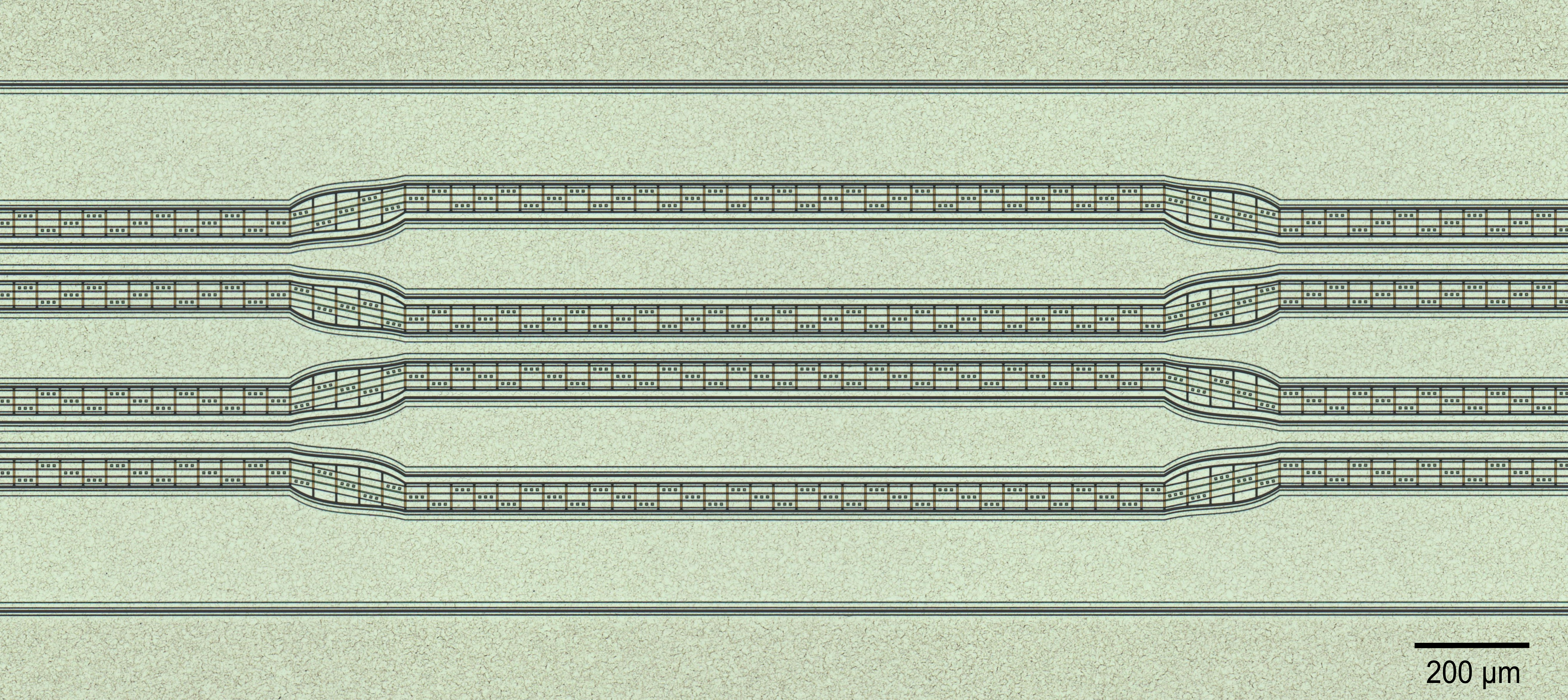}
    \caption{Optical microscopy image of the top-layer of the multi-layer version of the trap. The small black dots are the vias between the top and the bottom metal layer. They can be seen more clearly in figure~\ref{fig:MMT_tz}.}
    \label{fig:MMT}
\end{figure*}
\begin{figure}
    \centering
    \includegraphics[width=1\linewidth]{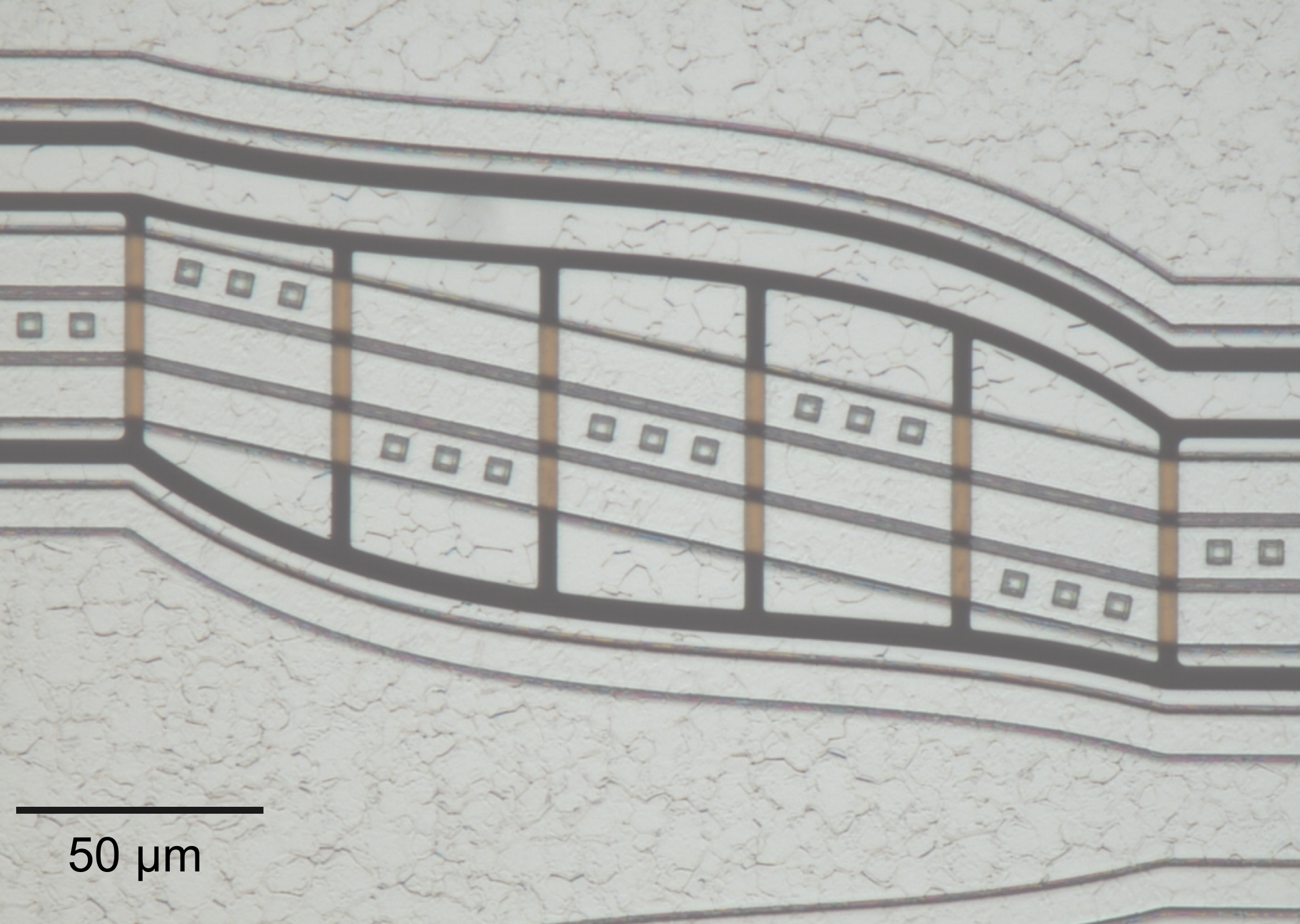}
    \caption{Optical microscopy image of a transition zone of the multi-layer version of the trap. One can see the topography from the routing lines in the metal layer below the top metal layer. The small black squares are the vias between the top and the bottom metal layer.}
    \label{fig:MMT_tz}
\end{figure}
The ion trap chip characterized in the previous section is a trap with a single structured metal layer. This allowed for fast prototyping but limits the number of dc electrodes, which can be independently routed, needed for finer control of multiple chains of ions manipulated simultaneously in a utility-scale trap. Therefore, we designed and fabricated a multi-layer version of the trap on fused silica substrate~\cite{Dietl2025} with two metal layers: one top layer and one bottom layer for routing of dc electrodes. The top layer consists of pure aluminum with a thickness of 2~µm, while the bottom layer has a thickness of 1~µm. Between them is an oxide layer ($\mathrm{SiO}_{x}$) of 1~µm thickness. The oxide layer was deposited through plasma-enhanced chemical vapor deposition (PECVD) and structured by plasma etching for vertical interconnect accesses (vias). The vias connect the two metal layers with each other. An image of the top layer of this trap is shown in figure~\ref{fig:MMT}. One can see the vias between the two layers as little black dots. The reliefs from the topography of the layer below can also be seen. The vias can be seen more clearly in the zoom-in on a transition zone in figure~\ref{fig:MMT_tz}. 

The rf electrode geometry is the same as in the single-layer version of the trap. The dc electrodes in the four lines between the rf electrodes consist of one compensation electrode in each line and rectangular dc electrodes. The dc electrodes in the four lines between the rf electrodes can be independently controlled. In the lines themselves, each third electrode is connected together, like for the trap in ref.~\cite{Holz2020}.  

The multi-layer stack will allow for the design and fabrication of even larger ion trap chips with more independently-controllable interaction and transition zones. With at least three metal layers, unshielded routing of dc connections below the rf electrodes can be avoided even for complex electrode geometries.

\section{Summary and conclusion}\label{sec:Conclusion}
In this paper, we have presented a trap concept enabling radial transport in a two-dimensional ion lattice, controlled by dc voltages. In the envisioned trap concept, one will be able to tune the radial coupling, and, most importantly, reach coupling strengths in the kHz-range, between ions in different wells of an rf double-well potential using only dc voltages. 

A design for an ion trap chip implementing a minimal instance was realized and tested. The layout of the rf electrodes is the result of an optimization procedure aiming at lowering the pseudo-potential barrier and ensuring only small variation of the total confinement in the transition zones. The final design results in a reduction of the pseudo-potential barrier to 3.8~meV, and a significant reduction of the variation of the secular frequency, compared to the unoptimized linear design, to below 1~MHz. The fabrication of an ion trap chip with optimized transition zones on a dielectric substrate using industrial CMOS fabrication processes has been executed and discussed. The functionality of the fabricated chip has been demonstrated and characterized. 

We have seen that it is possible to perform dc-controlled radial ion shuttling on an ion trap chip without ion loss at speeds in the range of a couple of mm/s. The possible speed should increase to 10 - 100 cm/s~\cite{Walther2012, Clark2023} by locally increasing the dc control in the transition zones by locally fine-tuning of the dc shim voltage set. Preciser stray field compensation at all shuttling positions reduces the influence of stray fields and reduces ion loss at higher shuttling speeds. The shuttling speed could be increased by replacing the RC filters with others having a higher cut-off frequency to allow for higher dc voltage update rates, while the addition of a low-noise amplifier can increase the available dc voltage range. 
The observed micromotion along the trap requires further investigation to verify that the micromotion can be reduced further.

Furthermore, the fabrication of a multi-layer chip was presented. The routing of an increased number of dc electrodes from the control electronics to the trap in the cryostat will require in the future cryogenic multiplexer chips~\cite{Malinowski2023, Brandl2017}. Higher electrode density will likely also make employment of through-substrate-vias necessary in future generations of ion trap chips. Furthermore, with an increased number of ions, it will become increasingly difficult to address all of the ions individually using free-space optics. More ions could be individually addressed by chiplets with photonic integrated circuits and microfabricated lens arrays that distribute the lasers through slots for optical access etched in the substrate~\cite{Badawi2025}. Another option would be the distribution of the light via waveguides and gratings. The exposed dielectrics from the integrated photonics can be shielded with transparent conductive oxides, such as indium tin oxide~\cite{Jansson2025}. The addition of slots for optical access will likely require to further complement the analytic design procedure for the interaction and transition zones by finite element methods~\cite{Oliveira2000, House2008}. 

\begin{acknowledgments}
We gratefully acknowledge support by the European Union’s Horizon Europe research and innovation program under Grant Agreement Number 101046968 (BRISQ).

We would like to thank all UPD and UPE engineers that supported us in the fabrication of the chips and all colleagues and friends in the quantum labs in Innsbruck and Villach that made the presented measurements possible.
\end{acknowledgments}

\bibliography{references}

\end{document}